\documentclass[superscriptaddress,twoside,twocolumn,showpacs,preprintnumbers,amssymb,pre]{revtex4}
\usepackage{graphicx}
\usepackage{dcolumn}
\usepackage{bm}
\begin{document}
\title{Stochastic Model for the Interaction of Buckling and 
       Fracture\\ in Thin Tension-Loaded Sheets}
\author{Bj{\o}rn Skjetne}
 \affiliation{Department of Chemical Engineering,
              Norwegian University of Science and Technology,\\
              N-7491 Trondheim, Norway}
 \affiliation{Department of Physics,
              Norwegian University of Science and Technology,\\
              N-7491 Trondheim, Norway}
\author{Torbj{\o}rn Helle}
 \affiliation{Department of Chemical Engineering,
              Norwegian University of Science and Technology,\\
              N-7491 Trondheim, Norway}
\author{Alex Hansen}
 \affiliation{Department of Physics,
              Norwegian University of Science and Technology,\\
              N-7491 Trondheim, Norway}
\date{\today}
\begin{abstract}
We introduce a model of fracture which includes the out-of-plane 
degrees of freedom necessary to describe buckling in a 
thin-sheet material. The model is a 
regular square lattice of elastic beams, rigidly connected at the 
nodes so as to preserve rotational invariance. Fracture is 
initiated by displacement control, applying a uniaxial force 
couple at the top and bottom rows of the lattice in mode-I type
loading. The approach lends itself naturally to the inclusion of
disorder and enables a wide variety of fracture behaviours
to be studied, ranging from systems with a simple geometrical 
discontinuity to more complex crack geometries and
random cracking. Breakdown can be initiated from a 
pre-cracked sheet or from an intact sheet where the first 
damage appears at random, and buckling sets in when a
displacement vector containing out-of-place components
becomes energetically favourable over one which does not. In 
this paper we only consider
center-cracked sheets with no disorder and include some results
relevant to the force- and displacement-fields, and
the buckling response ratio. Rather than 
carry out a comprehensive study of such systems, the emphasis 
presently is on the development of the model itself.
\end{abstract}
\pacs{81.40.Jj, 62.20.-x, 05.40.-a}
\maketitle

\section{Introduction}
\label{intro}
Understanding how, and when, materials break are important in 
many engineering applications. This is so for a number of 
reasons -- the motivation to study fracture may, for instance, 
be related to safety issues, such as determining when cracks 
form in concrete structures, or it may be one of economical 
gain, as in the case when the runnability of a printing press 
in the paper industry is considered. Few materials, be they 
natural or manufactured, are perfect, however. Hence, the
disorder in the micro-structure needs to be accounted for in
order to obtain a realistic description.

Over the past fifteen years methods 
have emerged within the statistical physics community to 
successfully tackle just such problems~\cite{smod}. 
These methods are intermediate between the microscopic, or
first principles, approach and the mean-field type of approach. 
In the former case fracture properties are derived from 
inter-molecular or inter-atomic forces, representing a problem 
which is both theoretically demanding and heavy on numerical 
resources. In the latter case disorder cannot be included in 
a satisfactory way. This is a big drawback since the presence 
of disorder in a material is crucial to the way it fractures. 
Disorder affects the stress field in such a way as to enhance 
the already existing heterogeneities. This interplay, between
a constantly evolving non-uniform stress field and local variations
in material properties, can nevertheless be handled in a
numerically tractable way using lattice models. 

The most common lattice models used in engineering applications
are finite element methods (FEM), the implementation of which
is usually based on commercially available computer codes.
The lattice models currently used in statistical physics differ
somewhat from the FEM-approach in that the grid used is 
regular, i.e., the same everywhere, rather than one which 
adjusts the mesh size according to where the stress field is 
most intense. Although FEM modeling is certainly more 
suitable in describing homogeneous materials, the requirement 
that the stress field should vary slowly over each element makes 
the approach cumbersome in the presence of heterogeneities. 
In the stochastic lattice model, however, the nodes are thought of 
as being connected by objects such as elastic beams or current 
carrying elements. While in some respects being less sophisticated 
than FEM methods, the interpretation of the algorithm is much 
more transparent and the approach also has the advantage of 
allowing disorder to be included quite generally. 

In the stochastic
models, the local equilibrium in force and moment is 
considered on a mesoscopic scale, i.e., on a scale much smaller 
than the external dimension of the lattice but still sufficiently 
large for the forces to be governed by well known physical laws. 
In this sense it is also a very good alternative to the far more 
complicated approach of including disorder on the microscopic level. 
Since only the nearest neighbours on the lattice 
are included the calculation of the displacement field reduces 
to the inversion a sparse matrix, enabling reasonably large 
systems to be handled computationally. 

One feature which is of a phenomenological nature, however, is 
the breaking rule -- the choice 
here is guided by intuition rather than by the inner workings 
of the model itself. In other words, breaking does 
not arise as a natural consequence of the calculations. This
is actually an advantage in that the mechanism by which the 
system ruptures can be tailored to suit different
engineering requirements. If we regard thin planar materials, 
for instance, the energy required to propagate a crack across a 
given area is usually much lower in mode-III fracture, i.e., 
tearing, than in the pure tensile loading of mode-I fracture. 
Familiar examples of disordered materials which behave this way are 
textiles and paper. 

Most of the research done so
far in stochastic lattice modeling aims to identify the
underlying general principles of the fracture process
rather than to address traditional problems in fracture
mechanics. In this paper the plane beam model~\cite{roux}, which
has been used previously to study scaling laws
in fracture, is extended to include a specific, and practical,
aspect of fracture which is very 
important for thin sheet materials, i.e., buckling.
As is well known, buckling can profoundly 
influence the residual strength of such 
materials~\cite{form,dixo,ziel}.
But before devoting our attention to this problem
in full, we briefly mention 
part of the background which has inspired the use of
lattice modeling as a tool in statistical physics.

In modeling experiments of random media, 
the feature which by far has received the most attention is 
the morphology of crack surfaces. 
Many surfaces in nature are found to be self-affine, i.e.,
statistically invariant with respect to anisotropic
scale transformations. The morphology of such surfaces
can be described by simple scaling laws, behaving very
much like fractal objects~\cite{mand}.
These scaling laws provide a theoretical framework whereby much 
information can be summed up in a few parameters.
Certain features have been found to share a common basis with other, 
seemingly unrelated, problems such as deposition and growth 
processes, or transport properties in random media~\cite{bara}. 
In the case of fracture it has been established that 
crack surfaces scale as $W\sim L^{\zeta}$,
where $L$ is the system size, $W$ is the roughness and
$\zeta$ is the roughness exponent. Other scaling laws have
been studied, e.g., in connection with the distribution of
stresses, or for the total amount of damage found at various 
stages in the breakdown process. 

By far the most popular
tool in such studies has been the random fuse model~\cite{fuse}.
In the fuse model,
the nodes on the lattice are connected by current-carrying
elements, i.e., fuses. The threshold for the amount of
current which may flow through each fuse is chosen 
from a random distribution. Hence, in the breakdown process a
fuse is irreversibly removed from the lattice once its threshold is
exceeded. A new distribution of currents is then 
calculated before the
next fuse is removed, and so on, until an uninterrupted path can
be traced across the system.
Although it really describes electrical breakdown, the
fuse model is often referred to as a scalar model of
fracture due to the similarity in form between Ohm's 
law and Hooke's law of linear elasticity. 
Results obtained for $\zeta$ with the fuse model are found to 
be different in two and three dimensions, however. The
results are $\zeta=0.74(2)$ in two dimensions~\cite{alex} and 
$\zeta=0.62(5)$ in three dimensions~\cite{batr}. Although
the former seems to agree with experimental findings,
the latter does not. Furthermore, the type of forces involved on 
the meso-scale also seem to make a difference, i.e., the
results obtained with a scalar model differ from those
obtained with a vectorial model. 
Specifically, in calculations with the elastic beam model
$\zeta=0.86(3)$ is obtained in two dimensions~\cite{skje}.
The difference between the results of
the two and three dimensional fuse model 
indicates that the additional degrees of freedom afforded 
by the (three-dimensional) buckling beam model should provide a 
lower estimate for the roughness exponent in the vectorial 
problem as well. Since the observed
value in real materials, i.e.,
$\zeta=0.8$~\cite{bouc}, in fact does lie below the two dimensional
beam lattice result, it would be interesting to see if
the buckling beam lattice
reproduces the universal value observed in nature.

However, although such fundamental aspects of the fracture process
are certainly interesting, the subject of how buckling affects
the scaling laws are left for future study.
The focus in this paper is instead on the development of a
lattice model that realistically includes the buckling 
behaviour observed in thin sheet materials. 
The characteristic out-of-plane deflection known as buckling 
is perhaps most frequently associated with thin plates or 
beams under compressive loading. 
Presently, however, we concern ourselves with the special case of a 
thin planar structure under tensile loading. The interaction 
of buckling with fracture in such circumstances is a well 
known phenomenon, although it has often been neglected in fracture 
mechanics analyses due to the extra complications involved.
One of the characteristic features of
buckling in a thin tension-loaded sheet is that a stable 
out-of-plane configuration is obtained after buckling has
set in. This is in stark contrast with the case of
compressive loading, where loss of stability usually signals
complete breakdown.

Practically all previous work considers the effect
buckling has on the strength properties of an already
cracked plate or a plate with a geometrical discontinuity 
such as a circular hole or a rectangular cut-out. If the 
physical parameters of the plate are such that buckling 
can be expected before the crack begins to grow, the
residual strength of the plate will be significantly
lower than what would otherwise be expected, based on an
analysis which does not take account of buckling. The
present study of fracture and buckling will be more general 
in scope. In other words, we also regard sheets which, in 
their initial state, have no cracks or other discontinuities.
Instead, cracks form by a complex process which depends 
on the evolving distribution of stresses 
and its interaction with a disordered meso-structure.
The onset of buckling in this scenario, and the effect
buckling has on the fracture properties, will vary according
to the type of disorder used, i.e., weak or strong. 
Nonetheless, even for weak disorders the final crack
which breaks the system will only rarely appear at the exact 
center of the sheet, and even then the situation will usually 
be complicated by additional cracks in the vicinity -- cracks 
which interact with the main crack so as to alter the 
distribution of stresses and hence also the exact shape or 
mode of buckling. 

The emphasis here, however, is on the development of the model 
itself. For illustration purposes, a few results are included 
on uniform systems with a center-crack. In section~\ref{planb}
the plane beam model is briefly reviewed, before the
equations describing the out-of-plane behaviour are 
derived in section~\ref{bbm}. Typical stress and
displacement fields are shown in sections~\ref{sdf} 
and~\ref{fff}, respectively, before the initialization of 
buckling is discussed in section~\ref{vrn} and a fracture 
criterion defined in section~\ref{fracrit}, 
where results for the buckling response ratio of
a center-cracked sheet are included. 
\begin{figure}
\includegraphics[angle=0,scale=0.65]{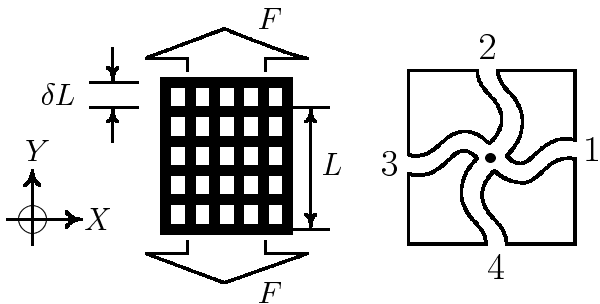}
\caption{On the left-hand side is shown a lattice of size $L=5$ 
         where a force couple has been applied uniformly on 
         opposite edges. The strain imposed is consistent with
         mode-I type fracture and corresponds to a displacement 
         $\delta L=1$ in the $Y$-direction. The enumeration 
         scheme of the neighbouring beams is shown on the 
         right-hand side, where a rotation at node~$i$ (center dot) induces 
         shearing forces and bending moments in the neighbouring 
         beams.
         \label{beamlatt}}
\end{figure}

\section{Plane Beam Lattice}
\label{planb}
The beam model may be defined as a 
regular square lattice of size $L\times L$, where the spacing 
is one unit length, and each node in the horizontal and vertical 
in-plane directions is connected to its nearest 
neighbours by elastic beams. A beam is then 
fastened to other beams in such a way that, upon 
subsequent displacement of neighbouring nodes, the angle 
between beams remains the same as in the original 
underlying square lattice, see Fig.~\ref{beamlatt}. 
Furthermore, all beams are imagined as having a certain 
thickness, providing finite shear elasticity. 

Beginning with the simple two dimensional beam model, 
there are three possible degrees 
of freedom, i.e., translations in the horizontal ($x$) and 
vertical ($y$) directions, and rotations about the axis 
perpendicular to the plane ($w$). As shown in
Fig.~\ref{beamlatt}, this allows for both 
bending moments and transverse shearing forces, in addition
to the axially tensile, or compressive, forces. 

For any node $i$, the nearest neighbours $j$ 
are numbered in an anti-clockwise manner, beginning with 
$j=1$ to the right of $i$. Defining $\delta r=r_{j}-r_{i}$,
where $r\in\{x,y,w\}$, the forces on $i$ due to $j=1$ 
are
\begin{eqnarray}
    _{w}M_{i}^{(1)}\hspace{-2mm}&=&\hspace{-2mm}
                \frac{1}{\beta+\frac{\gamma}{12}}
                 \bigl[\hspace{0.5mm}\frac{\beta}{\gamma}
                  \delta w
                  +\frac{\delta y}{2}
                   -\frac{1}{3}(w_{i}
                    +\frac{w_{j}}{2})
                      \hspace{0.5mm}\bigr],
                       \label{mi}\\
    _{y}T_{i}^{(1)}\hspace{-2mm}&=&\hspace{-2mm}
                \frac{1}{\beta+\frac{\gamma}{12}}
                 \bigl[\delta y-\frac{1}{2}\bigl(w_{i}+w_{j}
                  \bigr)\bigr],
                   \label{si}\\
    _{x}A_{i}^{(1)}\hspace{-2mm}&=&\hspace{-2mm}
                \frac{1}{\alpha}\delta x,
                 \label{fi}
\end{eqnarray}
for the moment due to angular displacements $w$, shear and 
transverse force due to displacements $y$, and axial strain 
due to displacements~$x$, respectively. Expressions for $j>1$ 
are analogous.

Prefactors characteristic of the material 
and its dimensions are
\begin{eqnarray}
    \alpha=\frac{1}{E\rho},\hspace{3mm}
     \beta=\frac{1}{G\rho},\hspace{3mm}
      \gamma=\frac{1}{EI},
       \label{mate}
\end{eqnarray}
where $E$ is Young's modulus, $\rho$ and $I$ the 
area of the beam section and its moment of inertia about 
the centroidal axis, respectively, and $G$ the shear 
modulus~\cite{herm}.

The conjugate gradient method~\cite{hest} is used to obtain 
the displacement field from
\begin{eqnarray}
    \sum_{j}D_{ij}
             \left[\begin{array}{l}
                      w_{i}\\
                      x_{i}\\
                      y_{i}\\
                    \end{array}
             \right]=\lambda
             \left[\begin{array}{l}
                      W_{i}\\
                      X_{i}\\
                      Y_{i}\\
                   \end{array}
             \right],
              \label{ma2x}
\end{eqnarray}
where 
\begin{eqnarray}
    X_{i}&=&{_{x}A_{i}^{(1)}}+{_{x}T_{i}^{(2)}}
         +{_{x}A_{i}^{(3)}}+{_{x}T_{i}^{(4)}},
         \label{sumfx}\\
    Y_{i}&=&{_{y}T_{i}^{(1)}}+{_{y}A_{i}^{(2)}}
         +{_{y}T_{i}^{(3)}}+{_{y}A_{i}^{(4)}},
         \label{sumfy}\\
    W_{i}&=&\sum_{j=1}^{4}{_{w}M_{i}^{(j)}},
         \label{sumfw}
\end{eqnarray}
are the components of force and moment. 

The material is assumed to be brittle, i.e., each beam
is linearly elastic up to the breaking threshold. Using 
$t_{A}$ and $t_{M}$ for the thresholds in axial strain
and bending moment, respectively, a good breaking 
criterion, inspired from Tresca's formula, is 
\begin{equation}
    \left(\frac{A}{t_{A}}\right)^{2}+
     \frac{|M|}{t_{M}}\geq 1,
      \label{tresca}
\end{equation} 
where $|M|={\rm max}(|M_{i}|,|M_{j}|)$ is the largest 
of the moments at the two beam ends $i$ and $j$. 

\begin{figure}
\includegraphics[angle=-90,scale=0.85]{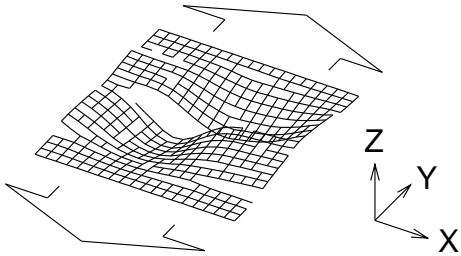}
\caption{A disordered beam lattice of size $L=19$, which is
         strained to failure in mode-I type fracture, i.e., by 
         applying a force couple at the top and bottom edges. 
         The presence of a central crack leads to the build-up
         of compressive stresses around the crack edges, causing 
         the structure to deflect out of the initial rest plane.
         \label{warp}}
\end{figure}

The fracture process is initiated by imposing 
an external vertical displacement which at the top row
corresponds to one beam-length, i.e., $\delta L=1$,
see Fig.~\ref{beamlatt}.
The lattice now consists of horizontally undeformed
beams and beams which in the vertical direction are 
stretched lengthwise.
The first beam to break is that for which the ratio
$A/t_{A}$ is largest, this being the vertically 
oriented beam which has the lowest value of $t_{A}$.
If all threshold values are the same, the 
next beam to break will be one of the nearest lateral 
neighbours since these now carry a larger load than 
other beams on the lattice. The case of no disorder is 
thus one in which the crack propagates horizontally 
from the initial damage, taking the shortest possible 
path to break the lattice apart. 
Introducing disorder in the breaking thresholds, material 
strength varies across the
lattice and consequently the crack will not necessarily 
develop from the initial damage point. Instead microcracks 
and voids form wherever the stress concentration
most exceeds the local strength, i.e., 
wherever Eq.~(\ref{tresca}) dictates that the next beam 
should be broken. Towards the end of the breakdown process 
smaller cracks merge into a macroscopic crack, forming a 
sinuous path which ultimately traverses the width of the lattice
and thus breaks it apart, see, for instance, Fig.~\ref{warp}.
In this scenario the quenched disorder on the thresholds
and the non-uniform stress distribution combine to determine 
where the next break will occur.
The stress distribution itself also continually changes as the 
damage spreads.

Throughout the process, the equilibrium stress field is
re-calculated by use of Eq.~(\ref{ma2x}) each time a beam 
is removed. The stress field therefore
relaxes at a rate much faster than the
process by which the crack grows. Hence, the model
describes quasi-static fracture.

\section{Buckling Beam Lattice}
\label{bbm}
The displacements of a real material, even if its geometry 
is essentially confined to a plane, will generally occupy
three dimensions. 
For instance, when opposite forces are applied uniformly 
along the top and bottom edges of a sheet of paper, with the 
object of straining it to failure, significant displacements 
will be observed in the direction perpendicular to the sheet,
see Fig.~\ref{warp}. 
This becomes especially evident wherever sizable cracks appear. 
Reasons for this behaviour are deviations 
in the symmetry of the material itself, or its properties, 
about the plane through which the externally applied forces 
act. In some cases such deviations may simply correspond to an 
uneven thickness, or they may be caused by local variations 
in density, a gradient in the orientation of the micro
structure, and so forth.

To include this behaviour, the plane beam model 
must incorporate at least two additional features. One is 
the random variation of the material in the out-of-plane 
direction. Since lattice modeling reduces the material to 
a set of points corresponding to the nodes on a mathematically 
precise two-dimensional lattice, the most convenient 
approach is to impose a very small randomly chosen vertical 
displacement on each node. This is discussed in more
detail in section~\ref{vrn}. The other feature to be 
included, and the topic of the present section, is the 
physics of the forces which create, and maintain, the 
out-of-plane displacement field.

In the buckling beam model we have one translational and one 
rotational displacement relevant to each of the principal 
axes, i.e., six degrees of freedom, with the matrix system
\begin{eqnarray}
    \sum_{j}D_{ij}
             \left[\begin{array}{l}
                      u_{i}\\
                      v_{i}\\
                      w_{i}\\
                      x_{i}\\
                      y_{i}\\
                      z_{i}\\
                    \end{array}
             \right]=\lambda
             \left[\begin{array}{l}
                      U_{i}\\
                      V_{i}\\
                      W_{i}\\
                      X_{i}\\
                      Y_{i}\\
                      Z_{i}\\
                   \end{array}
             \right]
              \label{ma3x}
\end{eqnarray}
replacing Eq.~(\ref{ma2x}). Presently the forces are 
projected onto the $XY$-, $XZ$ and $YZ$-planes, and hence
$X_{i}$ of Eq.~(\ref{ma3x}), that is, 
\begin{eqnarray}
    X_{i}&=&\sum_{j=1}^{4}X_{i}^{(j)},
             \label{sumx}
\end{eqnarray}
can be stated as
\begin{eqnarray}
     X_{i}&=&{_{x}^{\rm A}XX_{i}^{(1)}}
            +{_{x}^{\rm T}XY_{i}^{(1)}}
            +{_{x}^{\rm B}XY_{i}^{(1)}}
              \label{xf}\\
          &+&{_{x}^{\rm T}XZ_{i}^{(1)}}
            +{_{x}^{\rm B}XZ_{i}^{(1)}}
              \nonumber\\
          &+&{_{x}^{\rm A}XY_{i}^{(2)}}
            +{_{x}^{\rm T}XY_{i}^{(2)}}
            +{_{x}^{\rm B}XY_{i}^{(2)}}
              \nonumber\\
          &+&{_{x}^{\rm A}XX_{i}^{(3)}}
            +{_{x}^{\rm T}XY_{i}^{(3)}}
            +{_{x}^{\rm B}XY_{i}^{(3)}}
              \nonumber\\
          &+&{_{x}^{\rm T}XZ_{i}^{(3)}}
            +{_{x}^{\rm B}XZ_{i}^{(3)}}
              \nonumber\\
          &+&{_{x}^{\rm A}XY_{i}^{(4)}}
            +{_{x}^{\rm T}XY_{i}^{(4)}}
            +{_{x}^{\rm B}XY_{i}^{(4)}},
              \nonumber
\end{eqnarray}
where the term ${_{x}^{\rm B}XZ_{i}^{(1)}}$, for instance,
is the $x$-component of the buckling (B) force due to 
$j=1$, as projected onto the $XZ$-plane. Axial and transverse 
contributions are denoted (A) and (T), respectively.

The rotational displacements about the $Y$- and $X$-axes
are denoted $u$ and $v$, respectively, and $z$ is used for 
vertical displacements along the $Z$-axis. A coordinate 
system is placed 
on each node, whereupon forces and moments are expressed as
functions of the displacements. To this end, an elastic
beam with no end restraints~\cite{roar} is considered, as
in the case of the plane model. In the buckling model, however, 
the coordinate system is additionally rotated about the 
relevant angle within the $XZ$-, $YZ$- or $XY$-plane, 
i.e., $u$, $v$ or $w$.

With the exception of the signs on the terms of Eq.~(\ref{xf}), 
contributions from neighbours $j=1$ and $j=3$ are similar, 
as are those from $j=2$ and $j=4$. 

Consequently, if we define 
\begin{eqnarray}
    p_{j}=\frac{1}{2}\bigl[1-(-1)^{j}\bigr],
\end{eqnarray}
with $q_{j}=1-p_{j}$, and
\begin{eqnarray}
    r_{j}=\prod_{n=0}^{j-1}(-1)^{n}
\end{eqnarray}
with $s_{j}=(-1)^{j}r_{j}$,
for notational convenience, 
then the total force on $i$ along the $X$-axis, with the
contributions from all four of the neighbouring beams having been 
included, reads
\begin{figure}
\includegraphics[angle=0,scale=0.85]{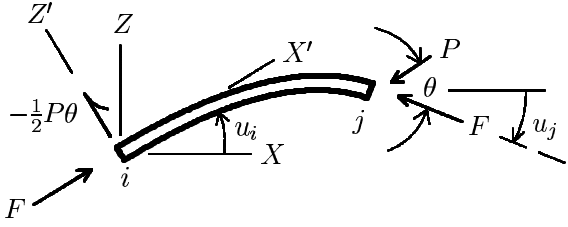}
\caption{The buckling term, $-\frac{1}{2}P\theta$, at node $i$ 
         due to $j=1$ in the case of an axially compressive load, 
         showing the angular displacements, $u_{i}$ and $u_{j}$, 
         the bending angle, $\theta$, the axial force, $F$, and
         the component $P$, of $F$, which is parallel to the beam 
         axis at node $i$. Also shown is the original $XZ$-system 
         and the X$^{\prime}$Z$^{\prime}$-system.
         \label{bucfig}}
\end{figure}
\begin{widetext}
\begin{eqnarray}
    X_{i}
      &=&
       \sum_{j=1}^{4}F_{i}^{(j)}
        \left[p_{j}\cos{w_{i}}
         \cos{u_{i}}- q_{j}\sin{w_{i}}\cos{v_{i}}-
          \frac{\delta w}{2}\cos{\delta w}
           \Bigl(p_{j}\sin{w_{i}}+ q_{j}\cos{w_{i}}\Bigr)
            \right.
             \label{xif}
              \\
      &-&\left.p_{j}\frac{\delta u}{2}\cos{\delta u}\sin{u_{i}}
        \right]s_{j}
         \nonumber\\
      &+&\frac{1}{\beta+\frac{\gamma_{\rm Z}}{12}}
          \sum_{j=1}^{4}
           \left[\bigl(p_{j}\delta x+q_{j}\delta y\bigr)
            \sin{w_{i}}+\bigl(-1\bigr)^{j}\bigl(p_{j}\delta y+q_{j}
             \delta x\bigr)\cos{w_{i}}-s_{j}\bigl(
              \frac{\delta w}{2}+\sin{w_{i}}\bigr)\right]
               \nonumber\\
      &\times&
        \Bigl(p_{j}\sin{w_{i}}+q_{j}\cos{w_{i}}\Bigr)
         \nonumber\\
      &+&\frac{1}{\beta+\frac{\gamma_{\rm Y}}{12}}
          \sum_{j=1}^{4}
           \left[\delta x\sin{u_{i}}+\bigl(-1\bigr)^{j}
            \delta z\cos{u_{i}}+r_{j}\bigl(\frac{\delta u}{2}+
             \sin{u_{i}}\bigr)
              \right]p_{j}\sin{u_{i}}.
               \nonumber
\end{eqnarray}

In Eq.~(\ref{xif}), moreover,
\begin{eqnarray}
    C(u,v)=p_{j}\frac{\delta u}{2}\bigl[sin(\frac{\delta u}{2})
                 \bigr]^{-1}
          +q_{j}\frac{\delta v}{2}\bigl[sin(\frac{\delta v}{2})
                 \bigr]^{-1}
                  \label{angc}
\end{eqnarray}
is an angular correction to
\begin{eqnarray}
    F_{i}^{(j)}=\frac{1}{\alpha}
       \left\{1-C(u,v)
        \sqrt{\delta z^{2}+
         \bigl[ 1-( p_{j}\delta x+q_{j}\delta y)s_{j}\bigr]^{2}}
          \hspace{1mm}\right\},
           \label{fiforce}
\end{eqnarray}
the latter being the projection onto the 
$XZ$-plane of the force 
along the axis of the beam. If we
consider the $j=1$ component in Eq.~(\ref{xf}), 
\begin{eqnarray}
    {_{x}^{\rm A}XX_{i}^{(1)}}=-F_{i}^{(1)}\cos w_{i}\cos u_{i}
         \label{axxi1}
\end{eqnarray}
is the contribution due to elongation or compression along the
axis of the beam,
\begin{eqnarray}
    {_{x}^{\rm T}XY_{i}^{(1)}}=
            \frac{1}{\beta+\frac{\gamma_{\rm Z}}{12}}
             \Bigl[\bigl(1+\delta x\bigr)\sin{w_{i}}-\delta y\cos{w_{i}}+
              \frac{\delta w}{2}\Bigr]\sin{w_{i}}
               \label{trxy}
\end{eqnarray}
is due to forces which are transverse to the axis 
of the beam, and
\begin{eqnarray}
    {_{x}^{\rm B}XY_{i}^{(1)}}=F_{i}^{(1)}
            \frac{\delta w}{2}\cos
             \delta w\sin w_{i}
              \label{buci1}
\end{eqnarray}
is the contribution due to buckling.
The latter arises when 
a beam in a bent configuration is under compressive or
tensile axial loading, see, e.g, Fig.~\ref{bucfig} where the term 
${_{z}^{\rm B}XZ_{i}^{(1)}}$ has been shown.
Out-of-plane contributions with components along the
$X$-axis are
\begin{eqnarray}
    {_{x}^{\rm T}XZ_{i}^{(1)}}=
            \frac{1}{\beta+\frac{\gamma_{\rm Y}}{12}}
             \Bigl[\bigl(1+\delta x\bigr)\sin{u_{i}}-\delta z\cos{u_{i}}+
              \frac{\delta u}{2}
               \Bigr]\sin{u_{i}}
                \label{trxz}
\end{eqnarray}
from transverse forces, and
\begin{eqnarray}
    {_{x}^{\rm B}XZ_{i}^{(1)}}=F_{i}^{(1)}
            \frac{\delta u}{2}\cos
             \delta u\sin u_{i}
              \label{bucxz}
\end{eqnarray}
\end{widetext}
from buckling.

The buckling term, as obtained in the lowest order 
approximation, is essentially the product of a bending angle, 
$\theta$, and an axial force component, $P$, the latter 
being parallel to the axis at the opposite end of the beam.
The component of the buckling reaction 
in the X$^{\prime}$Z$^{\prime}$-system which lies along
the $X$-axis in the $XZ$-system is then ${_{x}^{\rm B}XZ_{i}^{(1)}}$, 
i.e., Eq.~(\ref{bucxz}). 
 
Eq.~(\ref{trxz}), moreover, corresponds to
that part of the transverse force (including shear) 
which does not include buckling and is similarly 
obtained, i.e., by rotating the axes in Eq.~(\ref{si}). 

Finally, the axial term becomes
\begin{equation}
    \lim_{u,z\rightarrow0}{_{x}^{\rm A}XX_{i}^{(1)}}=
    {_{x}A_{i}^{(1)}}\cos w_{i}
    \label{limXX}
\end{equation}
when the out-of-plane displacements are set to zero. 
Hence, in this case, only when the rotation of the 
$XY$-system onto the X$^{\prime}$Y$^{\prime}$-system is 
neglected does Eq.~(\ref{xf}) reduce to Eq.~(\ref{fi}), 
of the plane beam model. 

Although forces and displacements on a beam under simultaneous 
axial and transverse loading cannot, in general, be obtained 
by superposition, combinations such as 
${_{x}^{\rm T}XZ_{i}^{(1)}}+{_{x}^{\rm B}XZ_{i}^{(1)}}$ in
Eq.~(\ref{xf}) result when only the leading terms, in $P$, are
retained after inverting the expressions of Ref.~\cite{roar}.
This also causes the buckling term in tensile loading to be 
the same as that in compressive loading, a change of sign 
being the only difference. 

The expression for the $Y$-component is similar
to that of the $X$-component, and is obtained by changing 
around the directions in Eq.~(\ref{xf}), i.e.,
\begin{eqnarray}
     Y_{i}&=&{_{y}^{\rm A}XY_{i}^{(1)}}
            +{_{y}^{\rm T}XY_{i}^{(1)}}
            +{_{y}^{\rm B}XY_{i}^{(1)}}
              \label{yf}\\
          &+&{_{y}^{\rm A}YY_{i}^{(2)}}
            +{_{y}^{\rm T}XY_{i}^{(2)}}
            +{_{y}^{\rm B}XY_{i}^{(2)}}
              \nonumber\\
          &+&{_{y}^{\rm T}YZ_{i}^{(2)}}
            +{_{y}^{\rm B}YZ_{i}^{(2)}}
              \nonumber\\
          &+&{_{y}^{\rm A}XY_{i}^{(3)}}
            +{_{y}^{\rm T}XY_{i}^{(3)}}
            +{_{y}^{\rm B}XY_{i}^{(3)}}
              \nonumber\\
          &+&{_{y}^{\rm A}YY_{i}^{(4)}}
            +{_{y}^{\rm T}XY_{i}^{(4)}}
            +{_{y}^{\rm B}XY_{i}^{(4)}}
              \nonumber\\
          &+&{_{y}^{\rm T}YZ_{i}^{(4)}}
            +{_{y}^{\rm B}YZ_{i}^{(4)}}.
              \nonumber
\end{eqnarray}
The $Z$-component, furthermore, is 
\begin{eqnarray}
     Z_{i}&=&{_{z}^{\rm A}XZ_{i}^{(1)}}
            +{_{z}^{\rm T}XZ_{i}^{(1)}}
            +{_{z}^{\rm B}XZ_{i}^{(1)}}
              \label{zf}\\
          &+&{_{z}^{\rm A}YZ_{i}^{(2)}}
            +{_{z}^{\rm T}YZ_{i}^{(2)}}
            +{_{z}^{\rm B}YZ_{i}^{(2)}}
              \nonumber\\
          &+&{_{z}^{\rm A}XZ_{i}^{(3)}}
            +{_{z}^{\rm T}XZ_{i}^{(3)}}
            +{_{z}^{\rm B}XZ_{i}^{(3)}}
              \nonumber\\
          &+&{_{z}^{\rm A}YZ_{i}^{(4)}}
            +{_{z}^{\rm T}YZ_{i}^{(4)}}
            +{_{z}^{\rm B}YZ_{i}^{(4)}},
              \nonumber
\end{eqnarray}
i.e., also similar in form to Eq.~(\ref{xf}) but 
with lines number two and five omitted. 

\begin{figure}
\includegraphics[angle=0,scale=0.75]{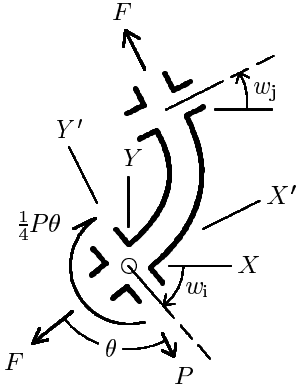}
\caption{The contribution $\frac{1}{4}P\theta$ to the in-plane 
         moment at node $i$ from $j=2$, due to buckling, in the 
         case of a tensile axial load. Shown are the angular 
         displacements, $w_{i}$ and $w_{j}$, the bending angle, 
         $\theta$, the axial force, $F$, and the component $P$, 
         of $F$, which is parallel to the axis of the beam at the
         opposite end, i.e., at node $j=2$. 
         \label{moment}}
\end{figure}

The full expressions are then
\begin{widetext}
\begin{eqnarray}
    Y_{i}
      &=&
       \sum_{j=1}^{4}F_{i}^{(j)}
        \left[q_{j}\cos{w_{i}}\cos{v_{i}}-p_{j}
         \sin{w_{i}}\cos{u_{i}}-
          \frac{\delta w}{2}\cos{\delta w}
           \Bigl(p_{j}\cos{w_{i}}+q_{j}\sin{w_{i}}\Bigr)
            \right.
             \label{yif}
              \\
      &-&\left.q_{j}\frac{\delta v}{2}\cos{\delta v}\sin{v_{i}}
        \right]r_{j}
         \nonumber\\
      &+&\frac{1}{\beta+\frac{\gamma_{\rm Z}}{12}}
          \sum_{j=1}^{4}
           \left[\bigl(-1\bigr)^{j}\bigl(p_{j}\delta x+q_{j}\delta y\bigr)
            \sin{w_{i}}+\bigl(p_{j}\delta y+q_{j}
             \delta x\bigr)\cos{w_{i}}-r_{j}\bigl(
              \frac{\delta w}{2}+\sin{w_{i}}\bigr)\right]
               \nonumber\\
      &\times&
        \Bigl(p_{j}\cos{w_{i}}+q_{j}\sin{w_{i}}\Bigr)
         \nonumber\\
      &+&\frac{1}{\beta+\frac{\gamma_{\rm X}}{12}}
          \sum_{j=1}^{4}
           \left[\bigl(-1\bigr)^{j}\delta y\sin{v_{i}}-
            \delta z\cos{v_{i}}-s_{j}\bigl(
             \frac{\delta v}{2}+\sin{v_{i}}\bigr)
              \right]q_{j}\sin{v_{i}}
               \nonumber
\end{eqnarray}
for the force in the horizontal direction, and
\begin{eqnarray}
    Z_{i}&=&-\sum_{j=1}^{4}F_{i}^{(j)}
             \left(p_{j}\sin{u_{i}}-q_{j}\sin{v_{i}}+p_{j}
              \frac{\delta u}{2}\cos{\delta u}\cos{u_{i}}-q_{j}
               \frac{\delta v}{2}\cos{\delta v}\cos{v_{i}}
                \right)r_{j},
                 \label{zif}
                  \\
         &-&\frac{1}{\beta+\frac{\gamma_{\rm Y}}{12}}
             \sum_{j=1}^{4}\left[\bigl(1+r_{j}\delta x\bigr)
              \sin{u_{i}}-r_{j}\delta z\cos{u_{i}}+\frac{\delta u}{2}
               \right]p_{j}\cos{u_{i}}
                \nonumber
                 \\
         &-&\frac{1}{\beta+\frac{\gamma_{\rm X}}{12}}
             \sum_{j=1}^{4}\left[\bigl(1-r_{j}\delta y\bigr)
              \sin{v_{i}}+r_{j}\delta z\cos{v_{i}}+\frac{\delta v}{2}
               \right]q_{j}\cos{v_{i}}
                \nonumber
\end{eqnarray}
for the force in the direction perpendicular to the rest plane.

Considering next the rotational contributions, a beam
under axial loading, which is simultaneously bent, is
shown in Fig.~\ref{moment}. In this case a buckling term 
arises which is again the product of a bending angle, 
$\theta$, and an axial force component. For rotations about 
the $Z$-axis, this gives
\begin{eqnarray}
    W_{i}&=&{_{w}^{\rm M}XY_{i}^{(1)}}
           +{_{w}^{\rm B}XY_{i}^{(1)}}
           +{_{w}^{\rm M}XY_{i}^{(2)}}
           +{_{w}^{\rm B}XY_{i}^{(2)}}
           +{_{w}^{\rm M}XY_{i}^{(3)}}
           +{_{w}^{\rm B}XY_{i}^{(3)}}
           +{_{w}^{\rm M}XY_{i}^{(4)}}
           +{_{w}^{\rm B}XY_{i}^{(4)}}.
            \label{wf}
\end{eqnarray}
\end{widetext}
For each beam in Eq.~(\ref{wf}) there are two terms, one 
analogous to Eq.~(\ref{mi}) and denoted (M), and one extra 
term, such as ${_{w}^{\rm B}XY_{i}^{(1)}}$, which is the
contribution due to the beam being simultaneously bent while 
under axial loading. Similarly, we have
\begin{eqnarray}
    U_{i}&=&{_{u}^{\rm M}XZ_{i}^{(1)}}
           +{_{u}^{\rm B}XZ_{i}^{(1)}}
           +{_{u}^{\rm Q}YY_{i}^{(2)}}
             \label{uf}\\
         &+&{_{u}^{\rm M}XZ_{i}^{(3)}}
           +{_{u}^{\rm B}XZ_{i}^{(3)}}
           +{_{u}^{\rm Q}YY_{i}^{(4)}},
             \nonumber
\end{eqnarray}
for rotations about the $Y$-axis, and
\begin{eqnarray}
    V_{i}&=&{_{v}^{\rm Q}XX_{i}^{(1)}}
           +{_{v}^{\rm M}YZ_{i}^{(2)}}
           +{_{v}^{\rm B}YZ_{i}^{(2)}}
             \label{vf}\\
         &+&{_{v}^{\rm Q}XX_{i}^{(3)}}
           +{_{v}^{\rm M}YZ_{i}^{(4)}}
           +{_{v}^{\rm B}YZ_{i}^{(4)}},
             \nonumber
\end{eqnarray}
for rotations about the $X$-axis, where (Q) denotes the
torque. In Eq.~(\ref{uf}), the torque is simply
\begin{eqnarray}
    {_{u}^{\rm Q}YY_{i}^{(2)}}=\xi\delta u,
            \label{torq}
\end{eqnarray}
where, assuming $w>t$, the material constant is
\begin{eqnarray}
    \xi=G\frac{wt^{3}}{3}
          \label{eqxi}
\end{eqnarray}
when $w$ denotes the width of the beam cross section
and $t$ its thickness. The buckling term reads
\begin{eqnarray}
    {_{u}^{\rm B}XZ_{i}^{(1)}}=-F_{i}^{(1)}
            \frac{\delta u}{4}\cos{\delta u},
             \label{momxz}
\end{eqnarray}
and that part of the bending moment which does not 
involve buckling becomes
\begin{widetext}
\begin{eqnarray}
    {_{u}^{\rm M}XZ_{i}^{(1)}}=
            \frac{\beta}{\gamma_{\rm Y}(\beta+\frac{
             \gamma_{\rm Y}}{12})}\delta u-\frac{1}{2(\beta+
              \frac{\gamma_{\rm Y}}{12})}
               \Bigl[\bigl(1+\delta u\bigr)\sin{u_{i}}-
                \delta z\cos{u_{i}}+
                 \frac{\delta u}{3}\Bigr]
                  \label{trmomxz}
\end{eqnarray}
when the axes are rotated.
Eq.~(\ref{uf}), when written out in full, is now
\begin{eqnarray}
    U_{i}=
      \frac{1}{\beta+\frac{\gamma_{\rm Y}}{12}}
       \sum_{j=1}^{4}p_{j}
        \left\{\frac{\beta}{\gamma_{\rm Y}}\delta u-\frac{1}{2}
         \Bigl[\bigl(1+r_{j}\delta x)\sin{u_{i}}-r_{j}
          \delta z\cos{u_{i}}+
           \frac{\delta u}{3}\Bigr]\right\}-
            \sum_{j=1}^{4}\left[F_{i}^{(j)}p_{j}
             \frac{\delta u}{4}\cos{\delta u}-q_{j}\xi\delta u\right]
              \label{uif}
\end{eqnarray}
and Eq.~(\ref{vf}) is analogous, i.e.,
\begin{eqnarray}
    V_{i}=
      \frac{1}{\beta+\frac{\gamma_{\rm X}}{12}}
       \sum_{j=1}^{4}q_{j}
        \left\{\frac{\beta}{\gamma_{\rm X}}\delta v-\frac{1}{2}
         \Bigl[\bigl(1-r_{j}\delta y)\sin{v_{i}}+r_{j}
          \delta z\cos{v_{i}}+
           \frac{\delta v}{3}\Bigr]\right\}-
            \sum_{j=1}^{4}\left[F_{i}^{(j)}q_{j}
             \frac{\delta v}{4}\cos{\delta v}-p_{j}\xi\delta v\right].
              \label{vif}
\end{eqnarray}
Finally, for rotations within the $XY$-plane, 
Eq.~(\ref{wf}) becomes
\begin{eqnarray}
    W_{i}&=&-\frac{1}{2(\beta+\frac{\gamma_{\rm Z}}{12})}
              \sum_{j=1}^{4}\left[
               \sin{w_{i}}-s_{j}
                \bigl(p_{j}\delta x+q_{j}\delta y\bigr)\sin{w_{i}}-r_{j}
                 \bigl(q_{j}\delta x+p_{j}\delta y\bigr)\cos{w_{i}}+
                  \frac{\delta w}{3}
                   \right]
                    \label{wif}
                     \\
         &+&\frac{\beta}{\gamma_{\rm Z}(\beta+\frac{\gamma_{\rm Z}}{12})}
             \sum_{j=1}^{4}\delta w-
              \frac{1}{4}\sum_{j=1}^{4}F_{i}^{(j)}\delta w\cos{w_{i}}.
               \nonumber
\end{eqnarray}
\end{widetext}

\begin{figure}
\includegraphics[angle=0,scale=0.70]{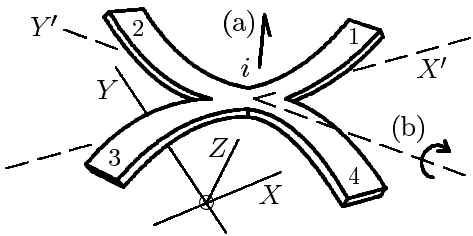}
\caption{Node $i$ and its nearest neighbours $j=1$--$4$, shown 
         when the lattice is in an advanced state of buckling. 
         The plane passing through $i$ is uniquely defined by 
         any $j$ and $j\pm1$ neighbours of $i$, as shown by the 
         broken lines, and is no longer parallel to the $XY$-plane. 
         The out-of-plane reaction (a) is normal to the 
         X$^{\prime}$Y$^{\prime}$-plane and (b) is a bending 
         moment about the Y$^{\prime}$-axis.
         \label{4beam}}
\end{figure}

In the six components of Eq.~(\ref{ma3x}), derived above, 
prefactors characteristic of the beam and its dimensions vary 
according to the principal axis of bending. Hence, in
Eq.~(\ref{mate}), we use
\begin{eqnarray}
    I_{\rm Z}=\frac{1}{12}w^{3}t
\end{eqnarray}
for bending within the $XY$-plane, and
\begin{eqnarray}
    I_{\rm X}=\frac{1}{12}wt^{3}=I_{\rm Y}
\end{eqnarray}
for bending within the $YZ$- and $XZ$-planes. We have then assumed 
beams with a rectangular cross-section, as already noted
in connection with Eq.~(\ref{eqxi}). This is convenient in
the study of how thin sheets behave during fracture, since
one may then simply visualize beams with a flat profile, see
Fig.~\ref{4beam}. In the present calculations the chosen
width-to-thickness ratio is 10:1, so that resistance towards
bending within the plane is much larger than that which 
governs out of plane bending.

In the following, results are displayed for non-disordered
systems with a central crack. To illustrate the nature of the
forces, moments and displacements involved, sections of the lattice
parallel with the crack are referred to as $J=1$, 2, ..., $L+2$.
Hence, on the bottom part of the lattice in Fig.~\ref{warp}, 
the set of nodes $i=1$ to $i=L+1$, located on the same row 
parallel with the $X$-axis, is referred to as $J=1$. With a total 
of $L+2$ rows $J$ parallel with the $Y$-axis, the ``near'' edge 
of the crack coincides with $J=L/2+1$ while the ``far'' edge 
coincides with $J=L/2+1$. Likewise, the set of nodes $i=1$ to
$i=L+2$ parallel with the $Y$-axis is referred to as, from left
to right, $I=1$ to $I=L+1$.

\section{Displacements}
\label{sdf}
The equations governing force and moment in a buckling beam 
lattice were derived in the previous section. At present we
have not taken account of the Poisson contraction which is 
observed in elastic systems -- at least not at the level 
of the individual beam. Such an effect does show
up, however, on length scales spanning several beams. Of 
course, in the macroscopic behaviour of the lattice, an 
example of this is precisely the buckling behaviour we intend to 
study. The bulging of the crack edges shown in Fig.~\ref{warp}, 
for instance, comes about as a result of transverse
compressive stresses which develop in the neighbourhood
of the crack. 

\begin{figure*}
\includegraphics[angle=-90,scale=0.32]{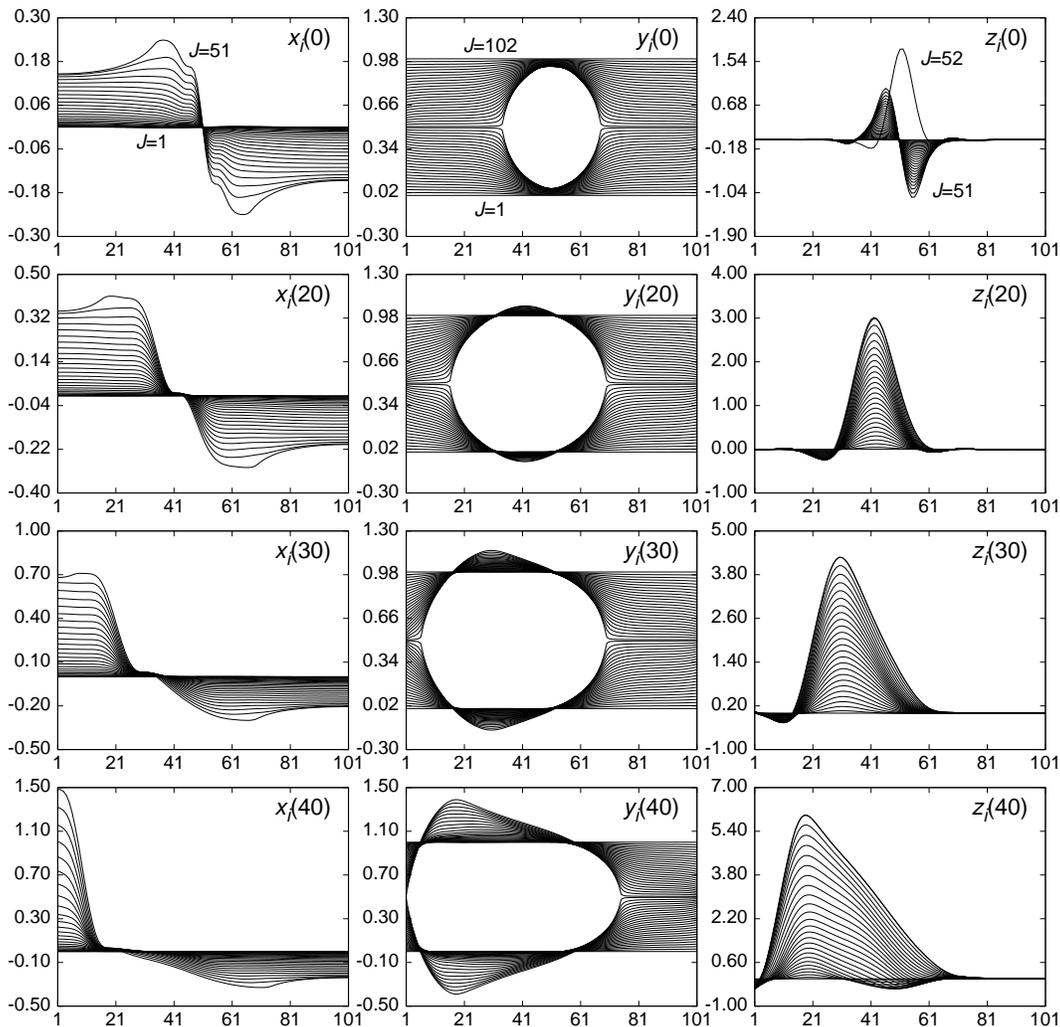}
\caption{Displacement fields across the width of a lattice of 
         size $L=100$ ($I=1$ to $I=101$) with an initial 
         center-crack. On the left-hand side $x_{i}(N)$ is shown 
         for $J=1$, 3, 5, etc., up to and including the crack interface, 
         i.e., $J=51$. At center is shown $y_{i}(N)$ for 
         $J=1$, 3, ..., 102. On the right-hand side, $z_{i}(N)$
         is shown for $J=1$, 3, ..., 51, including also the far side
         of the crack interface, i.e., $J=52$. The number of beams
         broken is $N$, and crack extent in the four stages shown
         is $I=34-68$ ($N=0$), $I=16-70$ ($N=20$), $I=6-70$ ($N=30$), 
         and $I=1-75$ ($N=40$).
         \label{3comp}}
\end{figure*}
Fig.~\ref{3comp} shows the in-plane displacements
$x_{i}$ and $y_{i}$ in a lattice of size $L=100$ at
various stages of crack advancement. The out-of-plane deflection 
$z_{i}(N)$ is also shown on the right-hand side. Displacements 
refer to the initial coordinate system on each node. In the
specific example shown the crack grows towards the left-hand
side of the lattice, with the out-of-plane deflection increasing
with the extent of the crack opening.
As previously mentioned, fracture is initialized by displacing 
the top row a unit distance. 
In the absence of geometrical discontinuities, each horizontal row 
$J$ is then incrementally displaced by an amount $(L+1)^{-1}$ 
with respect to the previous row $J-1$. In the absence of
cracks, the displacement field $y_{i}(J,N)$ then consists of a set
of equidistant lines between zero and one. With a crack present, 
this is altered into the pattern shown in Fig.~\ref{3comp}, 
e.g., for $y_{i}(0)$. As expected, transverse displacements 
$x_{i}(N)$ are largest close to the face of the crack. 
If we consider $x_{i}(0)$, and move from left to right along the
edge of the crack, beams are seen to be stretched wherever 
the slope is positive and compressed wherever it is 
negative. As the crack grows the net effect, however, is to cause 
the lattice to contract in the transverse direction, e.g., with 
the edge on the left-hand side moving inward by about 1.5 
beam lengths in the case of $x_{i}(40)$. 

The rotation of axes mentioned above is necessary 
to obtain the correct feedback between the force components
in the system, such as mutual consistency between
$XY$-forces and the $Z$-forces. To illustrate this, regard
the lattice before it begins to buckle, i.e., when the stress 
field is confined to the $XY$-plane. When a crack grows beyond 
a certain critical size, interaction between the stress 
field and random variations in the $Z$-direction initiates 
out-of-plane displacements which ultimately result in a 
buckled lattice. The driving forces are terms normal to the 
$XY$-plane, i.e., terms such as ${_{z}^{\rm B}XZ_{i}^{(1)}}$ and 
${_{z}^{\rm B}YZ_{i}^{(2)}}$, which belong to $Z_{i}^{(1)}$ 
and $Z_{i}^{(2)}$, respectively. These terms are not large.
In the flat lattice, for instance,
the last two terms on the right-hand side of
\begin{eqnarray}
     X_{i}^{(1)}
          &=&{_{x}^{\rm A}XX_{i}^{(1)}}
            +{_{x}^{\rm T}XY_{i}^{(1)}}
            +{_{x}^{\rm B}XY_{i}^{(1)}}
              \label{xf1}\\
          &+&{_{x}^{\rm T}XZ_{i}^{(1)}}
            +{_{x}^{\rm B}XZ_{i}^{(1)}}
              \nonumber
\end{eqnarray}
\begin{figure}
\includegraphics[angle=-90,scale=0.7]{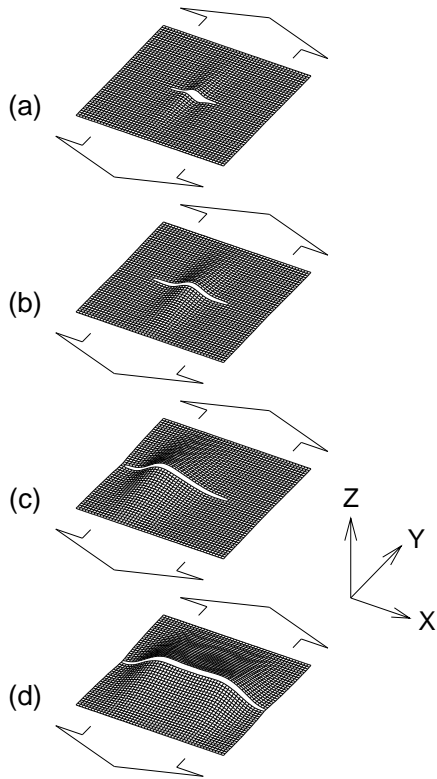}
\caption{A lattice of size $L=50$, with an initial center-crack,
         shown at four different stages of fracture. The number 
         beams broken are, from~(a) to~(d), $N=0$, 10, 20, and 34, 
         respectively.
         \label{w4}}
\end{figure}
\begin{figure}
\includegraphics[angle=-90,scale=0.248]{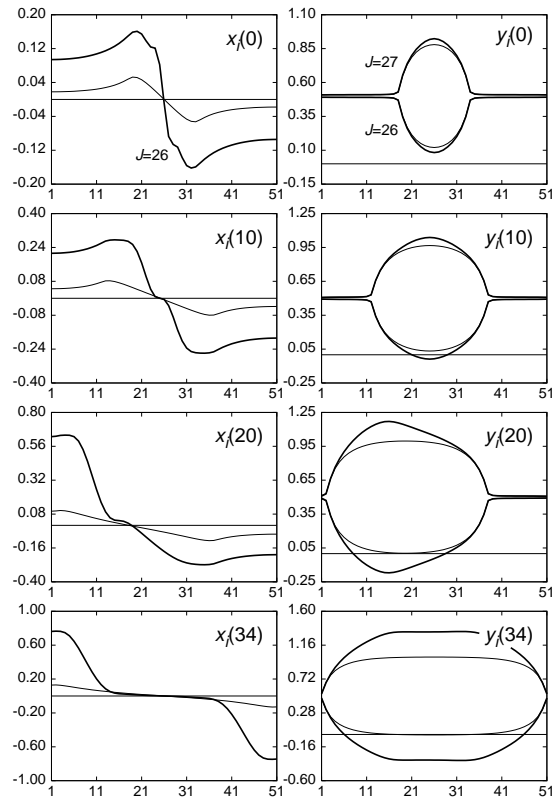}
\caption{Comparison between the crack-edge displacements, 
         obtained for the buckling lattice shown in Fig.~\ref{w4} 
         (thick lines) and the same lattice when the out-of-plane
         degrees of freedom are suppressed (thin lines). At the top, 
         the extent of the initial center-crack is $I=18-34$. 
         In subsequent stages, the crack extent is $I=12-38$ ($N=10$),
         $I=2-38$ ($N=20$) and $I=1-51$ ($N=34$). For $x_{i}(N)$ 
         the near edge of the crack is shown and for $y_{i}(N)$ both 
         edges are shown. 
         \label{p8}}
\end{figure}
are identically zero while the terms ${_{x}^{\rm T}XY_{i}^{(1)}}$ 
and ${_{x}^{\rm B}XY_{i}^{(1)}}$ are very small in 
comparison with the leading axial term. The terms 
${_{z}^{\rm B}XZ_{i}^{(1)}}$ and ${_{z}^{\rm B}YZ_{i}^{(2)}}$, 
however, although being small in comparison with 
either ${_{x}F_{i}^{(1)}}$ or ${_{x}T_{i}^{(2)}}$, are 
non-negligible. This owes to the fact that there is no 
physical obstruction in the lattice to inhibit displacements
in the $Z$-direction, i.e., there is no leading 
${_{z}^{\rm A}ZZ_{i}}$ term.
Moreover, as can be seen from Fig.~\ref{4beam}, when the 
lattice is in an advanced state of buckling there 
will be regions where the out-of-plane buckling reaction
is inclined with respect to the $Z$-axis, resulting in 
contributions of the type ${_{x}^{\rm B}XZ_{i}^{(1)}}\ne0$. 
These are smaller than ${_{z}^{\rm B}XZ_{i}^{(1)}}$,
but are also assumed to be non-negligible. In order to
include such terms the axes are rotated to
the local deflection of the lattice, whereupon the
components along the $X$-, $Y$- and $Z$-axes are obtained.

An example of the effect this has is when a large crack, 
perpendicular to the force couple, opens up in the center of 
the lattice, as in Fig.~\ref{warp} or Fig.~\ref{w4}. 
Although the initial 
displacement of the crack edges is normal to the rest plane, 
an in-plane component appears as the crack grows, i.e., the
near edge of the crack is pulled slightly back along the 
negative $Y$-axis while the far edge is pulled forward
in the opposite direction. This can be seen clearly in
Fig.~\ref{3comp}, where in $y_{i}(20)$ the rows nearest to the 
crack edges are displaced below or above the fixed values
of the top and bottom rows. In the case of the near edge
this means that displacements are negative, i.e., they have
moved slightly backwards with respect to their equilibrium 
positions in the unrestrained-strained lattice. These displacements 
become more pronounced as the crack grows, as can be seen 
from $y_{i}(30)$ and $y_{i}(40)$. 

It is particularly instructive to compare the displacements
of a buckling lattice with those obtained for the same lattice
when the out-of-plane degrees of freedom are suppressed.
In Fig.~\ref{w4} a lattice of size $L=50$ is shown in four
stages of crack advancement. The corresponding in-plane 
displacements of the crack-edges are shown as thick lines
in Fig.~\ref{p8}, with thin lines representing the same 
lattice in a non-buckling fracture mode. In the latter case, 
the $y_{i}$ displacements are seen to be confined between 
the fixed values of the top and bottom row, the physical 
structure of the lattice itself effectively acting as an 
obstruction to displacements outside this range. A further 
feature that can be noted concerns the aforementioned ``Poisson'' 
contraction. 
This effect is seen to be present in a 
non-buckling lattice as well, although to a much lesser 
degree. As for the angular displacements $w_{i}$ (not shown), 
these are seen to be somewhat larger in the buckling case, except 
for the peak values obtained at the crack tips, which are more 
or less the same. These peak values increase only as the crack 
nears the outer boundaries of the lattice, where $x_{i}$ is large.

In general, the expressions derived for force and moment 
from Ref.~\cite{roar} are accurate for small displacements only. 
This assumption we simply extend to all displacements. A 
second reason for rotating the axes, then, is to conserve 
the consistency of the approximations used. To illustrate 
the point one may, for instance, regard a straight beam 
which, in its rest state, lies along the $X$-axis. The term 
${_{z}^{\rm T}XZ_{i}^{(1)}}$ then expresses the transverse 
force as a function of $\delta z$. 
When $u_{i}$ is large, however, and the coordinate axes are
fixed, $\delta z$, in addition to the transverse force, 
also implies some measure of axial strain in the beam even 
though this effect has already been included via
Eq.~(\ref{fiforce}). Hence, rotating the axes precludes 
the introduction of systematic errors, e.g., due to angular 
deflections of the lattice. It also improves the quality
of the approximations used since angular displacements are 
rendered less severe in a rotated coordinate system. Large 
angular deflections usually involve a number of 
nodes on the lattice and hence by rotating the axes we avoid 
that too many errors accumulate.

It should also be pointed out, moreover, that
with three rotational degrees of freedom there will in 
principle be several displacement combinations 
which correspond to a given space orientation of the 
beam axis. Hence the projection of forces into the $XY$- 
$XZ$- and $YZ$-planes is an approximation based on the 
assumption that large deflections about more than one 
axis simultaneously are rare. Otherwise the exact orientation
of the beam would be history-dependent, i.e., it would
depend on the sequence in which the (final) angular 
displacements $u_{i}$, $v_{i}$ and $w_{i}$ were 
incremented.

\begin{figure}[t]
\includegraphics[angle=-90,scale=0.29]{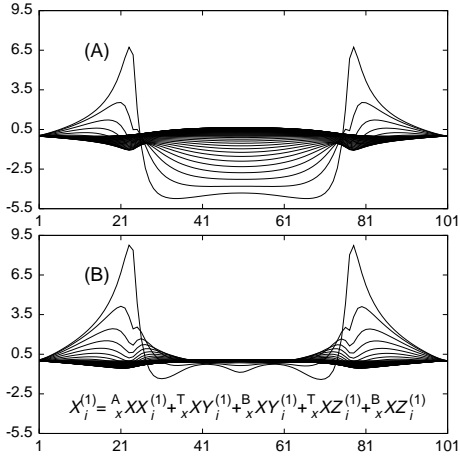}
\caption{Transverse stresses in a lattice of size $L=100$ with a 
         central crack which extends from $I=24$ to $I=78$. Shown 
         are stresses in the non-buckling (A) and buckling (B)
         fracture modes, for every row $J$ up to and including that 
         which coincides with the near edge of the crack, i.e.,
         $J=51$. Negative values correspond to compressive stresses.
         \label{twoX}}
\end{figure}
\section{Force Components}
\label{fff}
The out-of-plane force components are small, but their collective 
effect has a significant impact on the stress field. In the 
presence of significant cracks there is a feedback from the 
$Z$-displacements which allows the $XY$-displacements of the buckled 
lattice to relax with respect to the $XY$-displacements of the flat 
lattice. One example of this is the transverse compressive stress
stored in the region in front of and behind the crack in the
non-buckling lattice. Buckling releases this stress, as can
be seen in Fig.~\ref{twoX}, where the $j=1$ component of
Eq.~(\ref{sumx}) is shown in the non-buckling (A) and the
buckling (B) cases. In (A) a region of compressive stress confined 
between the crack tips is seen to extend for a distance of about 
6-8 rows away from the crack edge. In (B) only a vestige of
this is left, and then only in the immediate vicinity of the crack.
The tensile stress at the crack tips increases slightly 
in the buckled configuration.

In the non-buckling beam model, the extra non-linear terms
which arise when the beam is simultaneously bent while
under axial compression, or tension, are of lesser importance.
All forces now act within the structure which defines the 
plane so that, in calculating the in-plane displacement
field, corrections such as those due to $w_{i}$ in 
Eq.~(\ref{limXX}) may be neglected. The out-of-plane
displacement field, on the other hand, is obtained from an
equilibrium state in force and moment between a number of 
terms which are individually small. 

\begin{figure} [t]
\includegraphics[angle=-90,scale=0.32]{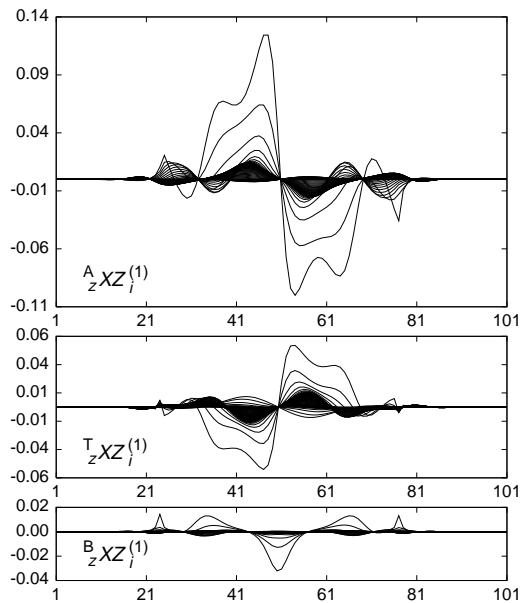}
\caption{Out-of-plane force components. Shown are the axial 
         ${_{z}^{\rm A}XZ_{i}^{(1)}}$, transverse 
         ${_{z}^{\rm T}XZ_{i}^{(1)}}$, and buckling 
         ${_{z}^{\rm B}XZ_{i}^{(1)}}$ contributions to $Z_{i}^{(1)}$ 
         of Eq.~(\ref{zf}). Contributions to $Z_{i}^{(3)}$ are
         similar, but with the contours of
         ${_{z}^{\rm A}XZ_{i}^{(3)}}$ and 
         ${_{z}^{\rm T}XZ_{i}^{(3)}}$ being mirror reflections
         of ${_{z}^{\rm A}XZ_{i}^{(1)}}$ and 
         ${_{z}^{\rm T}XZ_{i}^{(1)}}$ about $I=51$. 
         Contributions to $Z_{i}^{(2)}$ and $Z_{i}^{(4)}$ are smaller. 
         Lattice parameters used and contours shown
         are the same as in Fig.~\ref{twoX}.
         \label{compZ1}}
\end{figure}

For instance, the 
axial, transverse and buckling components which make up 
the $j=1$ contribution to Eq.~(\ref{zf}) are shown in 
Fig.~\ref{compZ1}. Here the vertical scales on the three 
subplots have been adjusted to the relative sizes of the 
components. The magnitudes, furthermore, refer to the scale 
of Fig.~\ref{twoX} and thus gives an idea of the ``smallness'' 
of the out-of-plane force components. 
In agreement with Fig.~\ref{twoX}, the forces in Fig.~\ref{compZ1} 
are seen to be most significant within a region nearest
to the crack edge, extending about 6-8 rows to either side
of the crack. 
\begin{figure} [b]
\includegraphics[angle=-90,scale=0.32]{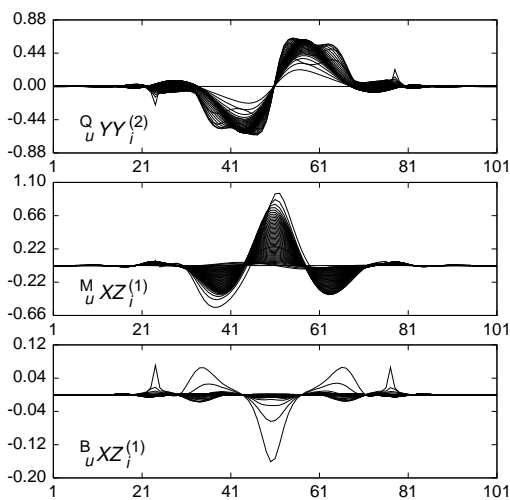}
\caption{Out-of-plane moment components about the $Y$-axis. Shown are the 
         torsional ${_{u}^{\rm Q}YY_{i}^{(2)}}$, axial 
         ${_{u}^{\rm M}XZ_{i}^{(1)}}$, and buckling
         ${_{u}^{\rm B}XZ_{i}^{(1)}}$ contributions to $U_{i}$ in 
         Eq.~(\ref{uf}). Lattice parameters used and contours
         shown are the same as in Fig.~\ref{twoX}.
         \label{Utor}}
\end{figure}
At the onset of buckling the axial and transverse terms,
${_{z}^{\rm A}XZ_{i}^{(1)}}$ and ${_{z}^{\rm T}XZ_{i}^{(1)}}$, 
are identically zero while the buckling term, 
${_{z}^{\rm B}XZ_{i}^{(1)}}$ shown in Fig.~\ref{bucfig},
is non-zero. In this-situation the sum of contributions from
$j=1-4$ is non-zero. As the out-of-plane deflection increases,
an equilibrium is approached where the sum of forces is
zero. At this equilibrium the buckling terms, e.g., terms such 
as ${_{z}^{\rm B}XZ_{i}^{(1)}}$ in Fig.~\ref{compZ1}, remain
non-zero. This is also the
case with the out-of-plane moments, shown in Fig.~\ref{Utor}.
The three subplots are not mutually to scale in
this case, but the magnitudes again refer to the scale of
Fig.~\ref{twoX}. Out-of-plane moments are seen to be 
somewhat larger than the axial and transverse buckling forces
of Fig.~\ref{compZ1}. Hence, at the equilibrium, the most significant of
the non-linear terms are those relevant to the momentum, 
${_{u}^{\rm B}XZ_{i}^{(1)}}$ being about five times larger than
${_{z}^{\rm B}XZ_{i}^{(1)}}$.

It is instructive to see what happens when buckling terms 
such as ${_{z}^{\rm B}XZ_{i}^{(1)}}$ and ${_{u}^{\rm B}XZ_{i}^{(1)}}$
are removed.
\begin{figure}
\includegraphics[angle=-90,scale=0.45]{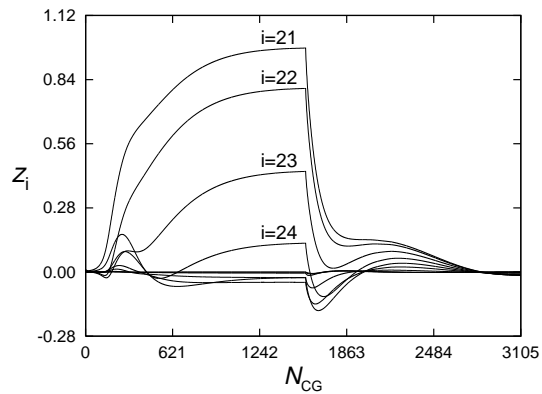}
\caption{The movement of the crack-edge as a function of the 
         time-steps in the numerical iteration, shown for a 
         lattice of size $L=40$ at the onset of buckling and for 
         a central crack between $I=14$ and $I=28$. Just prior 
         to  the point at which the equilibrium,  is reached, 
         at time $T_{\rm CG}=1570$, all terms such as 
         ${_{z}^{\rm B}XZ_{i}^{(1)}}$ are ``switched off''.
         \label{41CGR}}
\end{figure}
Shown in Fig.~\ref{41CGR}, at the onset of buckling, is the 
movement of the crack-edge as a function of time, the 
time-steps being defined by the iteration procedure which 
locates the equilibrium of force and moment. Just before
the point at which equilibrium is reached, all buckling terms
are ``switched off'' whereupon the remaining forces 
set about to locate a new minimum of elastic energy. This new 
minimum, of course, is none other than the flat configuration. 
As would be expected, not only do terms such as 
${_{z}^{\rm B}XZ_{i}^{(1)}}$ and ${_{u}^{\rm B}XZ_{i}^{(1)}}$ 
cause buckling, they also sustain it once it has been established.

\begin{figure*}
\includegraphics[angle=-90,scale=0.40]{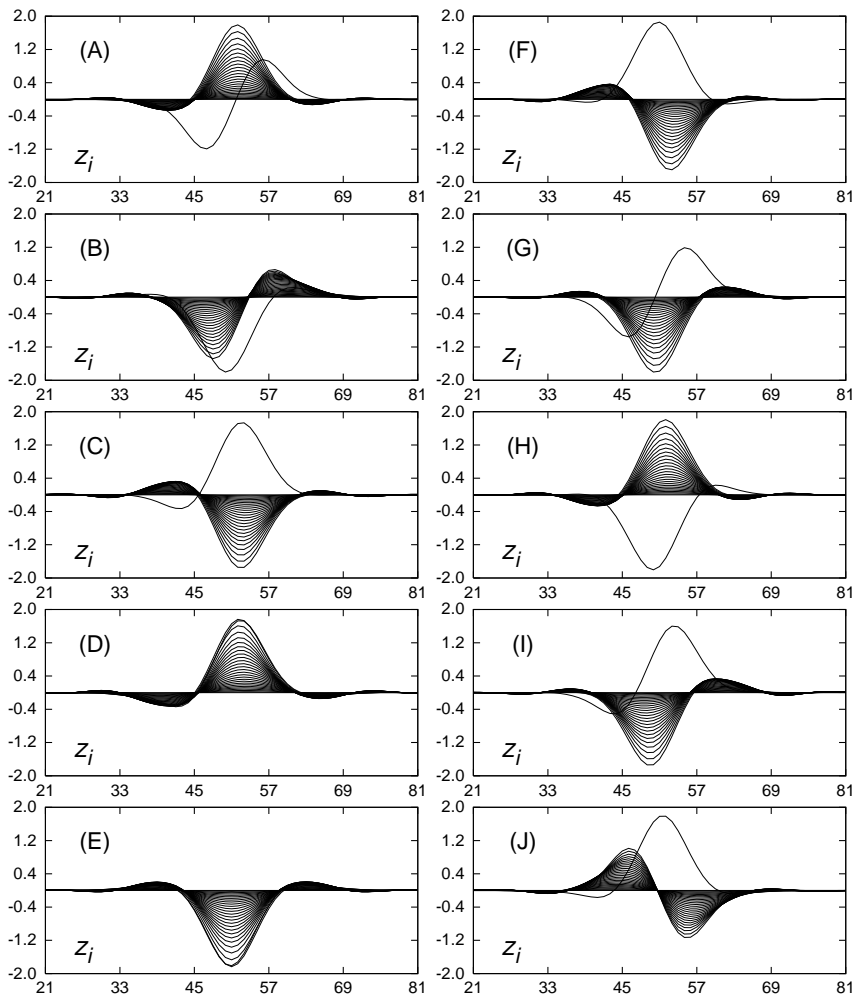}
\caption{Buckling modes for ten samples of a lattice of size $L=100$, 
         with a center-crack between $I=34$ and $I=68$. For the 
         lower half of the lattice, contours of every other row 
         $J=1$, 3, ..., are shown, up to and including $J=51$, i.e., 
         the near edge of the crack. The far edge, $J=52$, is also 
         shown. The only difference between the samples is the random 
         variation used to initialize the out-of-plane deflection. 
         \label{disp}}
\end{figure*}
Finally some remarks on the angular correction in 
Eq.~(\ref{fiforce}), which is included to allow for the
possibility that axial force may increase or decrease as a 
consequence of bending.  In Eq.~(\ref{fiforce}) 
it is assumed that the additional elongation due to bending 
can be obtained from a multiplicative factor. This factor
is based on the ratio of a circular arc~\cite{choi}
to a straight line, the former being the semi-circle defined by
the angular difference $\delta u$ at the end-points and the 
latter the line which connects these. On the level of the
individual beam, the presence of inflection points are 
neglected in this approximation. In other words, up-down 
curvatures may only occur in combinations of two or more 
beams in an end-to-end alignment. Furthermore, as can be 
seen from Eq.~(\ref{fiforce}), the effect of in-plane bending 
moments, or transverse displacements perpendicular to the 
rest axis of the beam, are neglected as contributions
which would otherwise add to the axial length of a beam.

\section{Initializing the Out-of-plane Deflection}
\label{vrn}
An important feature to be included in the model is 
the random variation of the material in the out-of-plane 
direction, as was remarked in section~\ref{bbm}. In thin 
materials such as paper, cloth, membranes and so forth, 
the most important factor influencing the behaviour 
during fracture is not the three-dimensional structure
of the material itself. Rather it is the out-of-plane deflection 
of this structure which makes a difference. Nevertheless,
random variations in the thickness direction provide an
important part of the mechanism which initiates buckling.
This is because such variations combine with the externally 
applied force and the emerging cracks to create local 
forces and moments which are not perfectly aligned within 
the plane. Once a buckled configuration has been established, 
however, the variation in the thickness direction is far
less important.

Since we presently regard the out-of-plane deflection of
a structure which has no vertical extent, buckling must be
initiated by other means. Specifically, in modeling the 
fracture process, the equilibrium stress field is 
re-calculated by use of Eq.~(\ref{ma3x}) after a beam 
has been removed. At each step of this process, i.e., for each
beam removed, a sample-specific random noise in the form 
of a small vertical displacement is imposed on all 
nodes of the lattice. Presently, we use a random number 
uniformly distributed on the interval $[-0.01,0.01]$.
In the early stages of the fracture process, the stress field is 
calculated in the presence of these variations 
until buckling commences. Before sizable 
cracks appear, forces combine to flatten out the vertical
displacements. That is, a flat configuration is energetically 
preferred to begin with, and fracture propagates according
to the non-buckling simulation. 
As significant cracks begin to appear, however, the 
conditions at some point become favourable for the out-of-plane 
components of the stress field to be realized, and buckling 
sets in. From here on the random noise is discarded, and 
the next displacement configuration is simply calculated from the 
previous coordinates. When the lattice has been broken, a 
new set of vertical displacements is generated for 
the next sample, i.e., a sample-specific random noise is used. 

Lattice buckling modes in the presence of a center-crack of 
size $\sim L/3$ are shown in Fig.~\ref{disp}.
The only disorder present here is that due to the out-of-plane
initialization, but evidently a number of buckling modes may
appear. 
In the following, cases where the deflection of the
edges of the crack is to the same side, i.e., up-up or down-down, 
are referred to as symmetric buckling, and cases where the
deflection is to opposite sides is referred to as anti-symmetric 
buckling. 
In Fig.~\ref{disp}, (D) and (E) are examples of the
former, and (F) and (H) are examples of the latter. 
Another buckling mode which frequently appears is that
shown in (A), (G) or (J), where the main bulge at one of the 
crack edges is made up of four, rather than three, half-waves. 
Based on the few dozen samples observed, (A) and (J) evolve into 
type (D) after a few more beams have been broken, and (G) into 
type (E), i.e., the symmetric buckling mode prevails in each
case. However, due to the randomness introduced, examples such 
as (H) are not completely anti-symmetric about the neutral plane.
Hence, even for this simple crack configuration, the exact shape 
of the out-of-plane deflection can vary considerably. The 
overall shape, however, tends to fall within the main 
categories, i.e., one of the two symmetric or anti-symmetric 
buckling modes. This way of initializing the out-of-plane
deflection is suitable for studying disordered systems,
where a lattice without any initial geometrical discontinuity
is strained until random cracks begin to appear. As the
cracks grow buckling sets in at some point, depending on
the configuration, position and size of the initial
cracks.

Another way of initializing the out-of-plane deflection is
to impose a small vertical deflection on a very few nodes in
strategic positions. This is most practical when studying
an ``ideal'' buckling scenario, such as the fracture of a 
non-disordered plate with a perfect center-crack. 
 
\begin{figure}[t]
\includegraphics[angle=-90,scale=0.40]{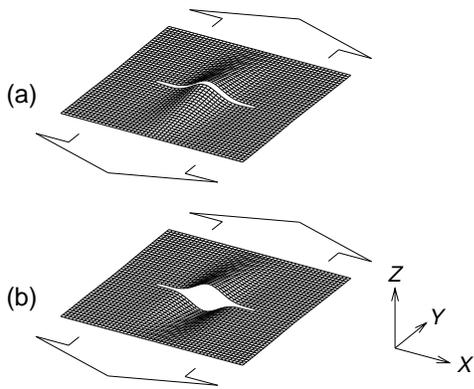}
\caption{A beam lattice of size $L=50$, showing (a) the symmetric
         and (b) the anti-symmetric buckling modes for a crack
         between $I=12$ and $I=40$.
         \label{2x50}}
\end{figure}
\section{Fracture Criterion}
\label{fracrit}
In order to study how buckling affects the fracture properties 
of a two-dimensional structure an appropriate breaking 
criterion should be chosen. As 
previously mentioned, this can be done to suit a range of
engineering requirements. Often the mode of rupture in the 
out-of-plane direction is radically different from that which
takes place within 
the plane. In paper or cloth, for instance, the 
phenomenon which first springs to mind is tearing. The energy 
required to propagate a crack across a given area in tear
mode is much less than that which causes the same area to 
fracture in pure tensile loading. This is especially the case
with paper.

Out-of-plane contributions to the breaking criterion must be
included by some other mechanism than that provided by
Eq.~(\ref{tresca}), since the latter is relevant to regions
which are comparable in size to a beam. The stress
intensification due to buckling, on the other hand, is
due to much smaller regions, i.e., comparable in extent
to the sharp crack tip. One way of enhancing the stress due
to buckling is to combine torsion with axial stress.
The larger the load, the more sensitive the beam will 
be to the presence of a given amount of torque. 
Compressive loads are assumed to alleviate the torsional
moment, but only to a very small degree. 

Hence, the breaking criterion can be stated as
\begin{equation}
    \left(\frac{F_{\rm C}}{t_{F_{\rm C}}}\right)^{2}+
     \frac{|\mu_{\rm C}|}{t_{\mu_{\rm C}}}\geq 1,
      \label{bcri}
\end{equation} 
where 
\begin{eqnarray}
    F_{\rm C}=F_{i}^{(j)}-\chi
        \Bigl|Q_{i}^{(j)}        
         \Bigr|                 
          \label{ft}
\end{eqnarray}
is the effective stress,
\begin{eqnarray}
    Q_{i}^{(j)}=p_{j}\cdot{_{v}^{\rm Q}XX_{i}^{(j)}}
               +q_{j}\cdot{_{u}^{\rm Q}YY_{i}^{(j)}}
                      \label{qq}
\end{eqnarray}
the torque, and $\mu_{\rm C}$ the combined bending moment.
With $w$ and $t$ denoting the width and thickness, respectively, 
of the beam, 
\begin{eqnarray}
    \sigma=\frac{w}{t}
\end{eqnarray}
is the aspect ratio of the cross section, and
\begin{equation}
    \chi=\left\{
         \begin{array}{cl}
            1+\sigma^{2}L\Bigl|F_{i}^{(j)}\Bigr|L_{0}, 
                    & \hspace{2mm}F_{i}^{(j)}<0,\\
            1,                                           
                    & \hspace{2mm}F_{i}^{(j)}\geq0,
         \end{array}\right.
          \label{tsens}
\end{equation}
is the enhancement factor in Eq.~(\ref{ft}).

Considering Fig.~\ref{2x50}, the breaking stress is increased
in case (b) and also in case (a) provided the deflections in
front of and behind the crack are not congruent. When the 
bulges are completely symmetric, however, (a) does not intensify 
the breaking stress and thus contradicts experimental findings.

Another possibility is to assume a crack-tip stress enhancement 
which depends on the out-of-plane bending moment. Here the in-plane 
displacement component $y_{i}$ observed in Fig.~\ref{p8}, i.e., 
the backward and forward movement of the crack edge, creates 
an angular displacement about the $X$-axis. For a sufficiently 
thin plate the resistance towards bending will not be
sufficient to halt the out-of-plane deflection once it has 
commenced, since the forces involved act over a region much 
larger than the immediate neighbourhood of the crack.
Due to the short distance which separates the top 
and bottom surfaces, the resulting ``lever-arm'' effect creates
an asymmetric stress-gradient across the crack front
in the direction of the thickness. Whereas tensile force on the 
concave side is then reduced, it increases on the convex side.
This increase comes in addition to the stress already
concentrated along the crack front, i.e., the very presence of a
crack creates a screening effect which re-distributes the
in-plane stresses so as to cause a build-up in the
load at the crack tips. For a crack that has grown to an extent
which allows buckling to occur, this in-plane stress
is significant. The crack-tip opening angle also plays an 
important role. Buckling in brittle materials, for instance, 
is known to have a profound effect on the maximum load
the system can tolerate before breaking. In 
a FEM study by Seshadri and Newman~\cite{sesh} a hypothetical 
very large 
critical crack-tip opening angle was used to model 
buckling in a ductile material. Strength 
reduction in this case was found to be significantly smaller 
than for brittle materials.

\begin{figure}
\includegraphics[angle=0,scale=0.38]{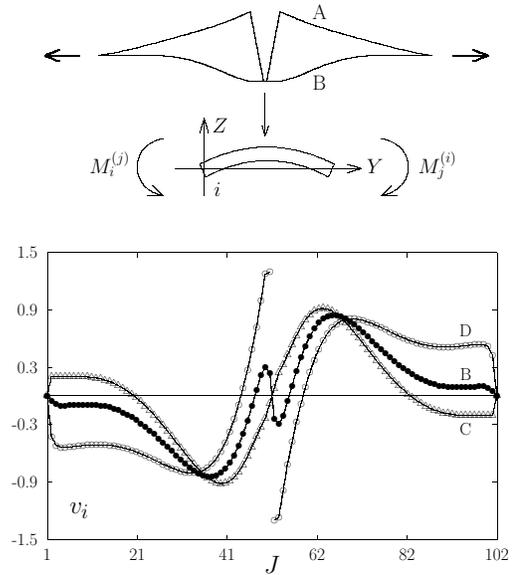}
\caption{Symmetric buckling. Shown at top, for a lattice of size
         $L=100$, is (A) the $I=51$ contour, passing through the 
         middle of the lattice, and (B) the $I=34$ contour, passing 
         through the left-hand side crack tip. Also shown is the
         bending mode of the beam which defines the crack tip at
         the junction between $I=34$ and $J=51$. At the bottom
         are shown the out-of-plane angular displacements $v_{i}$
         about the $X$-axis, in the case of (B) above. Also shown
         are the neighbouring contours, (C) $I=33$ and (D) $I=35$,
         where (D) is discontinuous due to the intersecting 
         crack.
         \label{uu}}
\end{figure}

\begin{figure}
\includegraphics[angle=0,scale=0.38]{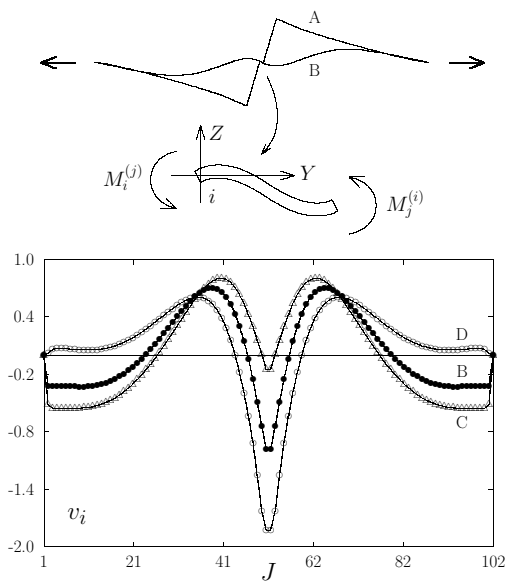}
\caption{Anti-symmetric buckling. Shown at top, for a lattice of size
         $L=100$, is (A) the $I=51$ contour, passing through the 
         middle of the lattice, and (B) the $I=34$ contour, passing 
         through the left-hand side crack tip. Also shown schematically
         is the the beam which defines the crack tip at
         the junction between $I=34$ and $J=51$. Although the
         bending mode is correct, the actual angles
         at the ends of the beam are negative and not positive 
         as shown. Out-of-plane angular displacements $v_{i}$
         about the $X$-axis are included at the bottom, with the
         notation being the same as in Fig.~\ref{uu}.
         \label{du}}
\end{figure}

In the beam model, the crack tip is never sharper than exactly 
one beam length. To emulate the above stress enhancement due 
to out-of-plane bending we instead imagine a sharp crack to be 
embedded within that beam which on the lattice defines the tip 
of the crack, and consider a combination of axial stress and 
moment. Out-of-plane bending modes are shown in 
Figs.~\ref{uu} and~\ref{du}, where the displacements of the 
schematic lattice at the top have been 
exaggerated somewhat to illustrate the point. Specifically,
the $Z$-displacements of contour B have been scaled up 100
times with respect to those of contour A, which itself is scaled
up with respect to the horizontal extent of the lattice.
The in-plane $Y$-displacements have also been adjusted 
accordingly.

Experimental evidence indicates that the stress enhancement 
at the crack tips is more or less 
similar in the symmetric and anti-symmetric buckling modes. 
To incorporate this we distinguish between the 
two cases. Hence, retaining Eq.~(\ref{bcri}), we introduce
\begin{eqnarray}
    \widetilde{M}_{i}^{(j)}=\frac{M_{i}^{(j)}}{|M_{i}^{(j)}|},
     \label{momdef}
\end{eqnarray}
where
\begin{eqnarray}
    M_{i}^{(j)}&=&p_{j}\left({_{u}^{\rm M}XZ_{i}^{(j)}}
                            +{_{u}^{\rm B}XZ_{i}^{(j)}}\right)
                                     \label{mm}\\
               &+&q_{j}\left({_{v}^{\rm M}YZ_{i}^{(j)}}
                            +{_{v}^{\rm B}YZ_{i}^{(j)}}\right)
                                     \nonumber
\end{eqnarray}
replaces $Q_{i}^{(j)}$ in Eq.~(\ref{ft}). In this prescription,
\begin{eqnarray}
    \widetilde{M}_{i}^{(j)}=\pm\widetilde{M}_{j}^{(i)}
\end{eqnarray}
denotes symmetric ($-$) or anti-symmetric ($+$) buckling,
respectively, with the signs referring to the direction of 
the moment at the two beam ends.

For the effective stress in the beam, we now use
\begin{equation}
    F_{\rm C}=F_{i}^{(j)}-\widehat{\chi}
        \Bigl|M_{i}^{(j)}-M_{j}^{(i)}\Bigr|
         \label{symmFc}
\end{equation}
in the symmetric case, i.e., when
\begin{equation}
    \widetilde{M}_{i}^{(j)}=-\widetilde{M}_{j}^{(i)},
\end{equation}
and
\begin{equation}
    F_{\rm C}=F_{i}^{(j)}-\frac{1}{2}\widehat{\chi}
        \hspace{1mm}{\rm max}(\Bigl|M_{i}^{(j)}\Bigr|,
         \Bigl|M_{j}^{(i)}\Bigr|)
          \label{antiFc}
\end{equation}
in the anti-symmetric case, i.e., when
\begin{equation}
    \widetilde{M}_{i}^{(j)}=\widetilde{M}_{j}^{(i)}.
\end{equation}
The enhancement factor in Eqs.~(\ref{symmFc}) and (\ref{antiFc}) is
\begin{equation}
    \widehat{\chi}=\left\{
         \begin{array}{cl}
            \widehat{\Lambda}_{i}
            \Bigl(1+\sigma^{2}L\Bigl|F_{i}^{(j)}\Bigr|L_{0}
            \Bigr), 
                    & \hspace{2mm}F_{i}^{(j)}<0,\\
            0,                                           
                    & \hspace{2mm}F_{i}^{(j)}\geq0,
         \end{array}\right.
          \label{bsens}
\end{equation}
where
\begin{eqnarray}
    \widehat{\Lambda}_{i}=p_{i}\widehat{\Lambda}_{i,x}
                         +q_{i}\widehat{\Lambda}_{i,y}
      \label{disco}
\end{eqnarray}
is a discontinuity operator.
The choice made above causes the breaking stress of the 
beams at the crack tips to increase by a comparable amount in
symmetric and anti-symmetric buckling.

The expressions for $\chi$ and $\widehat{\chi}$ in 
Eqs.~(\ref{tsens}) and~(\ref{bsens}), respectively,
have been chosen, very generally, to incorporate some overall
effects related to size, material and relative dimensions. 
Hence, it is reasonable to assume that, for a given 
size, ``tearability'', or the ``lever-arm'' effect, increases 
with decreasing sheet, or plate, thickness. 
In conjunction with 
this, the crack-tip opening angle, which decreases with increasing 
resistance towards in-plane bending, also enters the picture. 
Both effects are presently included via the ratio of
the in-plane to the out-of-plane inertial moment for bending, i.e.,
$\sigma^{2}=I_{\rm Z}/I_{\rm X}$.
The length of the arm with which the out-of-plane forces act is 
assumed to be proportional to the vertical extent of the buckling 
zone.  Since this, in turn, is proportional to system size, a 
factor $L\bigl|F_{i}^{(j)}\bigr|$ is also included.
As noted previously, fracture is initiated by 
displacing the top row of the lattice a fixed distance, usually 
corresponding to one beam length. To avoid scale effects associated 
with this, a further factor $L_{0}$ is included, where $L_{0}$
is the size of the reference system for which the top row
displacement is exactly one beam length. The introduction of a
reference system allows for the 
possibility of comparing systems of varying size where the 
physical behaviour involved requires the same relative external
boundary conditions. For instance, referring to the intact lattice, 
mode-I loading then imposes the same initial strain of 
$(L_{0}+1)^{-1}$ on each beam. 
Although computational time increases 
when $L>L_{0}$, features such as how various buckling modes
appear with respect to system size will depend on the external 
loading. Otherwise, $L_{0}=L$ might probably
be used in cases where we are interested in features which depend
on the internal processes of the fracture mechanism, such as the
roughness exponent of crack interfaces.

To guard against unphysical breaks, we introduce
\begin{eqnarray}
    \widehat{P}_{y,i}=n_{y,i-1}+n_{y,i+1}
\end{eqnarray}
which contributes when one or both nearest lateral neighbours are
intact, and
\begin{eqnarray}
     \widehat{Q}_{y,i}=\hspace{-1.5mm}
      \prod_{j=1}^{C_{L,m}-1}
       \hspace{-1mm}(1-n_{y,i-j})+\hspace{-2mm}
        \prod_{j=1}^{C_{L,m}-1}
         \hspace{-1mm}(1-n_{y,i+j})\hspace{0.5mm}
\end{eqnarray}
which
contributes when a certain number of neighbours have been broken. 
For any node $i$, the array
\begin{equation}
    n_{y,i}=\left\{
         \begin{array}{cc}
            0\\
            1
         \end{array}\right.
          \label{nindex}
\end{equation}
now keeps track of the status of the beam which extends away from
$i$ in the direction of the $Y$-axis, i.e., it remembers whether 
this is broken or intact, respectively. The combined
expression,
\begin{eqnarray}
    \widehat{\Lambda}_{y,i}=\widehat{P}_{y,i}\cdot
     \widehat{Q}_{y,i},
\end{eqnarray}
has the property
\begin{equation}
    \widehat{\Lambda}_{y,i}=\left\{
         \begin{array}{cc}
            0\\
            1
         \end{array}\right.
          \label{Lresu}
\end{equation}
as has $\Lambda_{x,i}$ governing cracks in the normal 
direction. In other words, Eq.~(\ref{disco}) ensures that the 
stress enhancement mechanism is activated only in cases where the 
lateral neighbour on one side is intact while simultaneously a 
certain number of beams, defining a minimum crack length 
$C_{L,m}$, are broken on the other side.

In most cases the operator $\Lambda_{i}$ is not necessary. It has
been included to avoid cracking being induced near the top and
bottom rows of the lattice. For very large systems, and especially
in cases where $L$ is significantly larger than $L_{0}$, breaks
sometimes occur due to the large angular gradients in beams
extending up from $J=1$ or down from $J=L+2$, see, for instance, 
Figs.~\ref{uu} and~\ref{du}.
In the present formalism the properties of beams with 
inflection points are not considered, the smallest crack that can
cause buckling, i.e., a bulge consisting of at least three 
half-waves, is therefore approximately $C_{L,m}=4$. This is
confirmed in numerical runs for systems with small $L$, but
where $L>>L_{0}$. 
A problem with using such a large value of $C_{L,m}$ is that 
it excludes cracks inclined at an angle with respect to the 
horizontal. Over a wide range of system parameters and external 
boundary conditions, however, $C_{L,m}=2$ was found to be 
adequate. 

\begin{figure} [t]
\includegraphics[angle=-90,scale=0.40]{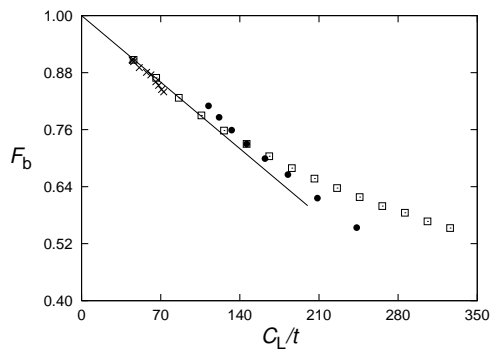}
\caption{The buckling response ratio, $F_{b}$, shown as a function 
         of the crack-length-to-thickness ratio, $C_{L}/t$, for
         various systems with $L=3C_{L}$. Open squares denote the
         results of varying $L$ while keeping $t$ constant, filled 
         circles denote the results of varying $t$ with $L=54$ 
         fixed, and crosses denote similar results with $L=24$ 
         fixed. Kuhn and Figge's linear expression~\cite{kuhn} is 
         also included for comparison.
         \label{kuhn}}
\end{figure}
In the limit of no buckling, i.e., $\delta v\rightarrow 0$
or $\delta u\rightarrow 0$, Eq.~(\ref{bcri}) reduces to 
Eq.~(\ref{tresca}). In other words, if buckling is not activated, 
then neither is the stress enhancement mechanism
in the fracture criterion.

Finally, in order to illustrate how Eq.~(\ref{bcri}), with
Eqs.~(\ref{momdef}) to~(\ref{Lresu}) defining the stress 
enhancement mechanism, works within our model of buckling, we
consider the buckling response ratio of the residual strength
of the system. That is, $F_{\rm b}=\lambda_{0}/\lambda_{Z}$,
where $\lambda_{0}$ and $\lambda_{z}$ represent the maximum 
applied external force a restrained or buckled plate, respectively, 
can tolerate before breaking apart. Early experimental
results show that the decrease in strength due to buckling 
increases as the ratio of the crack-length $C_{L}$ to the 
thickness $t$ is increased. A linear relationship was
proposed by Kuhn and Figge~\cite{kuhn} which, in the case of 
brittle materials, has been shown to agree 
well with more recent FEM calculations~\cite{sesh}. 
In Fig.~\ref{kuhn}, results obtained with the beam model are
compared with the Kuhn-Figge relationship. A small correction to
the size of the central crack has been made to account for the 
finite size of the beams. The effect is very small, shifting
the values of the smallest systems slightly to the left, thus
improving the agreement with the Kuhn-Figge relationship from 
very good to excellent.

\section{Summary}
\label{ssm}
To summarize, we have included the additional degrees of freedom
necessary to describe the interaction of cracks with buckling
in the elastic beam model. This model is stochastic in nature,
so that sheets with random cracking at any level of 
meso-structural disorder can be studied, including systems
with no disorder.
In addition to important issues of practical relevance
in traditional 
fracture mechanics, such as strength properties and stability,
the present model also enables fundamental aspects of
fracture in random media to be explored.

\end{document}